# Flexible and Actuating Nanoporous Poly(ionic liquids)-paper based Hybrid Membranes

*Huijuan Lin,[†] Jiang Gong,[†] Han Miao,[†] Ryan Guterman,[†] Haojie Song,[‡] Qiang Zhao,[†] John W. C. Dunlop[§,]\* and Jiayin Yuan[†,]\**

[†]Department of Colloid Chemistry, Max Planck Institute of Colloids and Interfaces,

Am Mühlenberg 1 OT Golm, D-14476 Potsdam, Germany

[‡]School of Materials Science and Engineering, Jiangsu University,

Zhenjiang, Jiangsu, 212013, China

[§]Department of Biomaterials, Max Planck Institute of Colloids and Interfaces,

Am Mühlenberg 1 OT Golm, D-14476 Potsdam, Germany

ABSTRACT: Porous and flexible actuating materials are important in the development of smart systems. We report here a facile method to prepare scalable, flexible actuating porous membranes based on a poly(ionic liquid)-modified tissue paper. The targeted membrane property profile was based on a synergy of a gradient porous structure of poly(ionic liquid) network and the flexibility of tissue paper. The gradient porous structure was built up through ammonia-triggered electrostatic complexation of a poly(ionic liquid) with poly(acrylic acid) (PAA) that were previously impregnated inside the tissue paper. As a result, these porous membranes undergo bending deformation in response to organic solvents in vapor or liquid phase and can recover their shape back in air, which was demonstrated to be able to serve as solvent sensors. Besides, they show enhanced mechanical properties due to the introduction of mechanically flexible tissue paper that allows the membranes to be designed as new responsive textiles and contractile actuators.

## 1. INTRODUCTION

Increasing attention has been recently paid to "smart" materials that can undergo shape deformation or color variations in response to external environmental stimuli.[1-18] Among these smart materials, flexible polymer-based actuators have been realized for many promising applications including solvent sensors, humidity monitoring, soft robotic, *etc*.[19-22] Many actuators function through the development of an asymmetric, unbalanced distribution of internal strains inside an actuator.[23,24] There have been numerous actuator designs over the past decade in literature, which from a structure point of view include bilayers,[25] trilayers,[26] gradient types,[27] gels,[28] and so forth. Currently, the structural design of actuators is under expansion, and new actuation concepts are eagerly pursued. Among these attempts, an emerging trend is to introduce porous networks into the responsive matrix because pores can add extra merits such as light-weight, possible variation of optic properties, as well as accelerated mass transport and diffusion rate inside the actuator. For example, a porous, three-layer polyimide film actuator was fabricated by Ionov *et al.*, of which the inhomogeneous pore density is created by etching out silica particles. The actuation mechanism is purely based on the asymmetric distribution of the pores inside a polyamide film that varies the swellability of the top and bottom zone of the film in contact with solvent.[29] Very recently, our group has reported, *via* electrostatic complexation between a cationic poly(ionic liquid) (PIL) and multi-acid, to prepare a series of (nano)porous PIL-derived membranes, which feature particularly a gradient of crosslinking density along the membrane cross-section from the top to the bottom. The co-existence of pores and the structural gradient enables membranes to bend quickly in contact with solvents or their vapors due to different swelling degrees inside the membrane.[27,30] Unfortunately, due to a restricted mechanical property window of PIL membranes, it is still mechanically insufficient to be incorporated into flexible smart devices and difficult to achieve complex folding for more sophisticated applications.

To realize flexible porous actuators, several materials have been applied as flexible substrates such as textiles,[31] plastics,[32] papers.[33-37] Among these substrates, papers have been widely utilized to construct multifunctional materials for various applications[38-50] for its low cost, ultra-lightweight, high tensile strength, mechanical flexibility, biocompatibility, and addressable functionality.

To explore porous actuators for more complex motions such as elongation/contraction instead of a simple bending that is often used as a basic actuation model for mechanistic study, new material designs with high flexibility are urgently needed. Besides, extension of the response to other stimuli for the cellulose paper-based actuators is highly demanded but remains a challenge for practical applications. To tackle these challenges, herein a hybrid nanoporous membrane was fabricated from commercial cellulose tissue paper, whose interior was filled up with nanoporous PILs. This hybrid membrane carries mechanical properties similar to the native tissue paper, while keeping the nanoporous character and the gradient feature from the PIL porous network to exert actuation function in response to various organic solvents and their vapors. Derived from their flexible nature, examples of complex designs, such as responsive textile and contractive actuators were fabricated to demonstrate a high level control of physical motions of soft actuators. This work points out a new general synthetic approach to prepare biomass-derived functional actuators.

## 2. EXPERIMENTAL SECTION

**2.1 Materials.** Lithium bis(trifluoromethane sulfonyl)imide (LiTf$_2$N, 99.95%), poly(acrylic acid) (PAA, $M_w$ = 1800 g/mol), and aqueous ammonia (28 wt.%) were purchased from Sigma-Aldrich and used without further purification. Poly(3-cyanomethy-l-vinylimidazolium bromide) (PCMVImBr) was synthesized according to our previous report.[51] The white tissue papers were obtained from KIMTECH SCIENCE with a single sheet size of 11.2 cm × 21.3 cm. The solvents used were of analytic grade.

**2.2 Preparation of PIL-PAA@tissue paper membranes.** Poly[3-cyanomethyl-1-vinylimidazolium bis(trifluoromethane sulfonyl)imide] (PCMVImTf$_2$N, simplified as "PIL") was synthesized by anion exchange from PCMVImBr with LiTf$_2$N. For membrane preparation, PIL (1.00 g) and PAA (0.18 g) were dissolved in 10 mL N,N-dimethylformamide (DMF) in a 1:1 molar ratio of monomer unit. A piece of pristine paper with a size of 2 cm × 5 cm was put onto a polytetrafluoroethylene (PTFE) substrate and coated with 1 mL of PIL-PAA/DMF solution. After dried at 80 °C for 3 h and soaked in 0.2 wt.% aqueous ammonia for 2 h, the composite membrane, denoted as PIL-PAA@tissue paper, was finally removed from the substrate, washed with water several times, and stored in wet condition before use.

**2.3 Characterization.** The structures and morphologies of pristine papers and the composite membranes were characterized by scanning electron microscopy (SEM, GEMINI LEO 1550 microscope, 3 kV). Attenuated total reflection Fourier transformed infra-red (ATR-FTIR) measurements were recorded on a Nicolet™iS5 instrument from Thermo Scientific. Mechanical properties of the dry membranes were measured on an Instron 1121 at an extension speed of 10 mm/min. Nitrogen sorption measurement was performed using a Quantachrome Autosorb-1C-MS analyzer at 77 K. The hybrid membranes were degassed at 100 °C for 20 h before the measurements. The mercury porosimetry experiments were conduct to measure the porosity and pore size distribution of pristine papers and hybrid membranes with an Autopore III device (Micromeritics, USA) according to DIN 66133.

## 3. RESULTS AND DISCUSSION

The fabrication procedure of the hybrid membrane is shown in Figure 1a. Generally speaking, it takes the advantages of our previously used approach to create the porous structure and the gradient profile along the membrane cross-section.[30] Here, a hydrophobic cationic PIL, poly[3-cyanomethyl-1-vinylimidazolium bis(trifluoromethane sulfony)imide], was mixed with PAA ($M_w$ = 1800 g/mol) in DMF in a 1:1 molar ratio with regard to the repeating unit to form a uniform mixture solution. PAA is known to exist in a protonated thus noncharged form in DMF as a polar aprotic solvent, which is homogeneously mixed with the cationic PIL in DMF on a molecular level. The mixture solution was then cast onto, instead of a glass plate used in our previous method, a piece of pristine tissue paper (2 cm × 5 cm × 60 μm) lying flat on a PTFE substrate. The tissue paper employed in this work is composed of an interpenetrating network of cellulose fibers of about 20 μm in diameter, which are visualized by scanning electron microscopy (SEM) in Figure 1b and Figure S1. The nonuniformly packed fibers create abundant interconnected interstices that can accommodate the to-be-introduced porous PIL network.

After being coated with the polymer blend solution, the paper lying on the PTFE substrate was dried at 80 °C for 3 h to evaporate DMF, leaving a dry blend film sticking firmly to the PTFE substrate. Here, the film surface in direct contact with the PTFE substrate is termed as BOTTOM surface, while the other side towards air as TOP surface. The dry film together with the PTFE substrate was soaked in an aqueous ammonia (0.2 wt. %) solution for 2 h to develop the targeted porous hybrid membrane, denoted as PIL-PAA@tissue paper, which was easily peeled off from the substrate after the soaking treatment. Due to the facile fabrication process, a large-sized membrane can be easily obtained, *e.g.*, 480 cm$^2$ (Figure S2) that was synthesized in our lab. It should be noted that the ammonia treatment is a crucial step, during which ammonia and water molecules diffused into the film.[52] The former neutralized PAA in the blend film to in-situ electrostatically crosslink the PIL chains *via* interpolyelectrolyte complexation, while the latter introduced pores *via* phase separation of the hydrophobic nature of the PIL in water because of a high content of bis(trifluoromethane sulfony)imide) (Tf$_2$N$^-$), a large-sized hydrophobic fluorinated anion. Essentially, the pores were generated *via* a concurrent phase-separation/crosslinking process. During this process, a gradient in the crosslinking degree from top to bottom along the ammonia diffusion track was built up. As each ionic crosslinking point is formed by the release of a Tf$_2$N$^-$ anion which is the only sulfur-containing specie, the sulfur content along the cross-section of the membrane (top: 6.05 wt. %, bottom: 9.85 wt. %) is a nature outcome and directly related to the crosslinking degree gradient (Figure S3). By analyzing the Fourier transform infrared spectroscopy (FT-IR) spectra of the dry film before and after the ammonia treatment, we observed that the absorption band at 1700 cm$^{-1}$ that is assigned to C=O stretching in the COOH form reduced its intensity significantly, while the band at 1550 cm$^{-1}$ which is ascribed to the carbonyl stretching mode in COO⁻NH$_4^+$ is dramatically enhanced, confirming the occurrence of the deprotonation of carboxylates during the ionic complexation process (Figure S4).

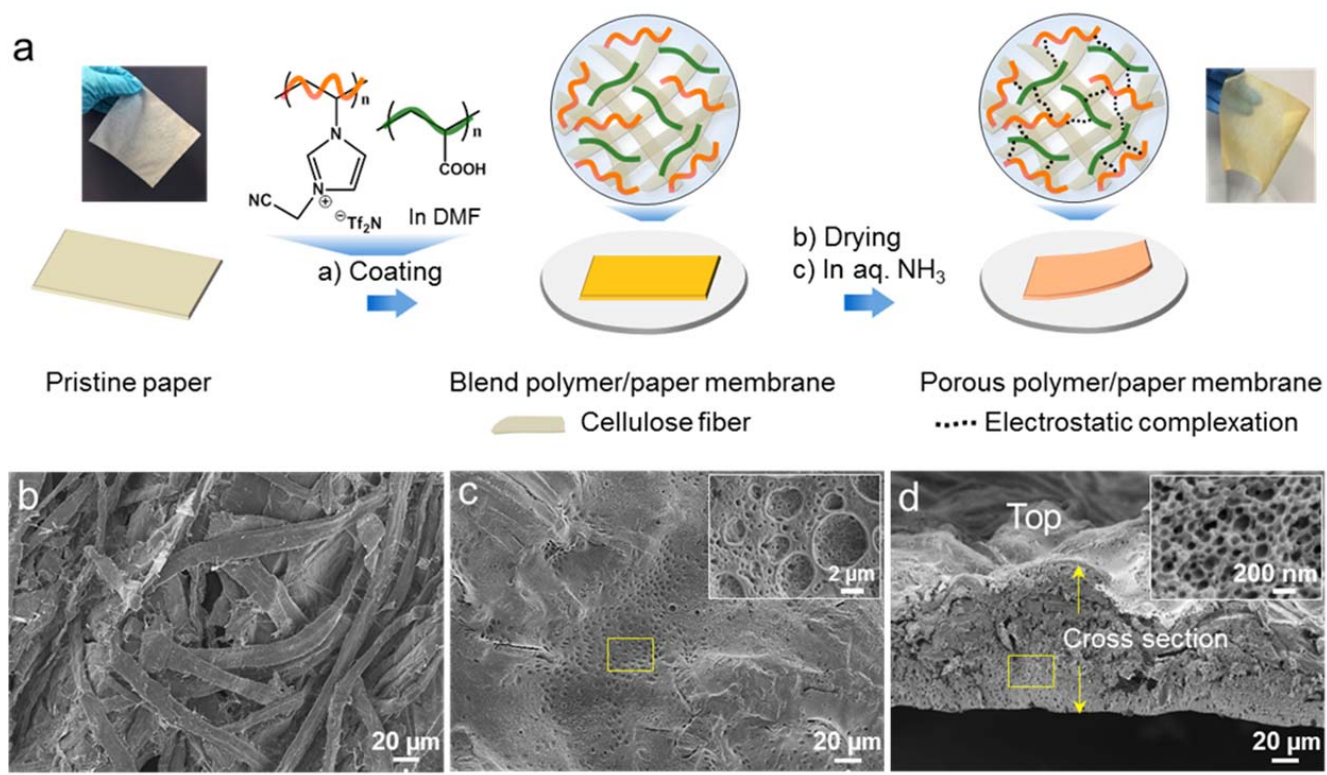

**Figure 1.** (a) Schematic illustration of the fabrication procedure of the hybrid PIL-PAA@tissue paper membrane. The photographic insets are the pristine tissue paper (left side) and the hybrid PIL-PAA@tissue paper membrane (right side). (b) SEM image of the top view of the tissue paper. (c) SEM image of the top view of the hybrid membrane. Inset is a SEM image of the region marked with a yellow box in (c). (d) SEM image of the cross-section view of the hybrid membrane. Inset is the SEM image of the region marked with a yellow box in (d).

The pristine tissue paper (inset on the left side in Figure 1a) is white in color and randomly foldable at will. The inset on the right side in Figure 1a is the as-fabricated hybrid PIL-PAA@tissue paper membrane. Except being pale yellowish, it resembles the original paper. Both top and bottom surfaces appear homogenous by naked eyes. The bottom surface is indeed flat when further visualized by SEM (Figure S5), while the top (Figure 1c) is densely packed with submicron sized pores that allows for liquid diffusion. After the formation of the hybrid membrane, the fibrous morphology in the pristine paper is no longer visible due to the filling-up by the porous polymer in the interstice among cellulose fibers inside the paper. The inner structure of the hybrid membrane is shown in Figure 1d as a cross-section view of the membrane. Compared with that of the pristine tissue paper (Figure S1) that is 60 µm thick, the cross-section in the hybrid membrane increases its thickness to 100 µm, which is in fact tunable by coating different volumes of the polymer solution (Figure S6) into the paper matrix, that is, the more polymers are added, the thicker the final hybrid membrane will be. This dimensional expansion was previously observed in a tissue paper-free native PIL-PAA porous membrane because of the pore volume introduced during the ammonia treatment of the thinner PILs/acid blend film. In a close view of the cross-section in Figure 1d, it is seen that the cross-section is not fully filled with polymer, and micron-scale voids are present. Even by adding more polymer solution to fill the interstice during the membrane fabrication process, these voids remain in the hybrid membrane, implying that these voids are more likely intrinsically generated during the ammonia treatment, since the PIL-PAA/tissue paper blend membrane shows no pores or voids before ammonia treatment (Figure S7). We assume that the expansion in the membrane thickness is responsible for the void formation, as the polymer matrix and the fibrous network have different expansion co-efficiency, *i.e.* they expand at different scales. Finally, a high-resolution view of the cross-section (inset in Figure 1d) shows the dense nanoporous texture. The pore size is in the range of 50-100 nm, which is identical to that of a native PIL-PAA porous membrane when produced in the absence of the tissue paper (Figure S8).

To investigate the porosity and pore size distribution of the native tissue paper and the PIL-PAA@tissue paper membrane, mercury porosimetry measurements were conducted. As shown in Figure 2a, the tissue paper has a large mercury uptake of 4.5 mL/g at a relatively low pressure (below 1 MPa) and nearly no uptake above 10 MPa, indicating a significant amount of interstice among cellulose fibers with pore size in micron range (average pore size = 60 μm in Figure 2b). This large interstice and high porosity (89%) in tissue paper is important to maintain the porous structure of the introduced PIL-PAA component. In the case of PIL-PAA@tissue paper, the mercury uptake was significantly reduced to 0.5 mL/g even at 100 MPa (porosity of 46%), indicating the interstice among the cellulose fibers is now filled by porous PIL-PAA. In its pore size distribution, the dominant pore size at 60 μm vanished due to polymer filling, and a weak broad peak from several to tens of nm appeared at a pressure above 50 MPa, corresponding to the introduced porous polymer network. Nitrogen sorption measurement of the PIL-PAA@tissue paper (Figure S9) reveals a non-defined isotherm and a Brunauer–Emmett–Teller specific surface area of 2 $m^2$/g, *i.e.* it is neither a micro- or mesoporous structure as expected.

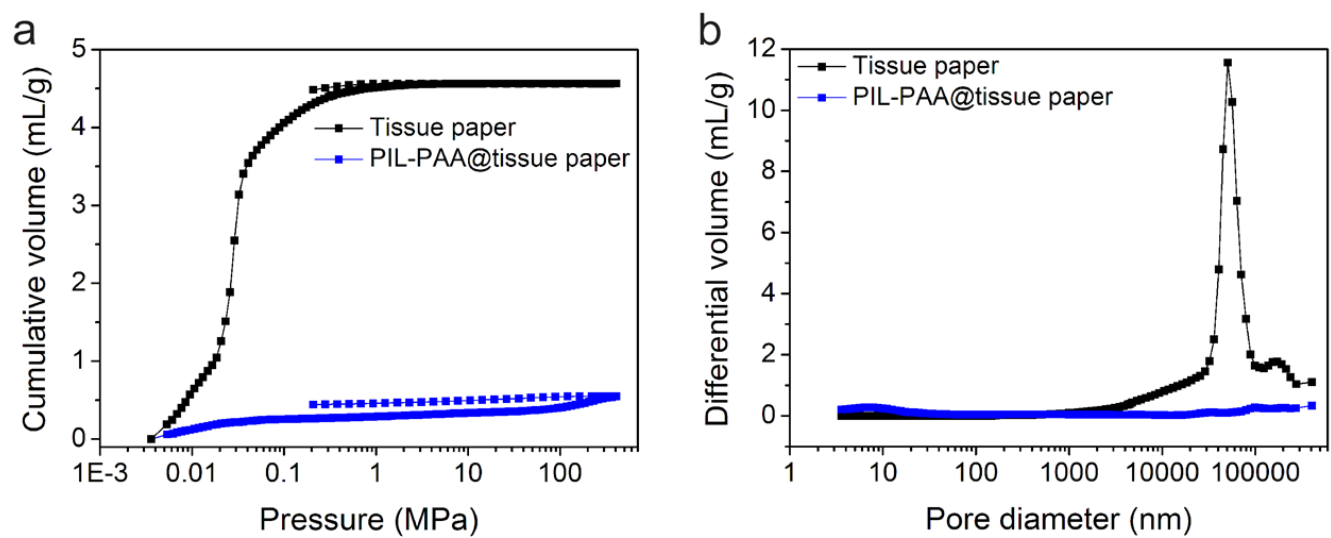

**Figure 2**. Mercury intrusion measurements of the tissue paper and the PIL-PAA@tissue paper membrane. (a) Cumulative mercury volume absorbed by tissue paper and PIL-PAA@tissue paper membrane as a function of the pressure. (b) The differential pore volume of tissue paper and PIL-PAA@tissue paper membrane as a function of the pore size.

Favorably the hybrid membrane in a dry state is easily folded into complex patterns, a character that is carried by the pristine tissue paper and now passed to the hybrid membrane. As an example, the membrane was readily folded or processed into an artificial flower (inset in Figure 3), indicating its excellent mechanical flexibility and processability. By contrast, the native PIL-PAA membrane in a dry state cracks into small pieces upon mechanical bending. This comparison justifies the motivation of applying the tissue paper as scaffold for the porous polymer membrane to overcome its mechanical dilemma. To quantitatively analyze the mechanical properties, the hybrid membrane was measured together with the pristine paper and the native PIL-PAA membrane. As shown in Figure 3, the native PIL-PAA membrane breaks at a low strain of *ca.* 0.1%, which is much lower than the strain at break for the other two samples, being 2.5% of the hybrid membrane and 6% of the pristine tissue paper. As for the tensile strength at failure, the hybrid membrane presents a value of 3.0 MPa, being not only significantly higher than 1.7 MPa of the native PIL-PAA membrane but also surprisingly surpassing 2.7 MPa of the pristine tissue paper, which is indicative of a synergistic interaction between the porous PIL network and the cellulose fibers. From the well-recognized knowledge on PILs, the polyvinylimidazolium PILs are able to bind various surfaces ranging from metals to cellulose.[53] In the current case, the porous network serves also as a binder to glue individual cellulose fibers together into a network, thus enhancing its tensile strength.

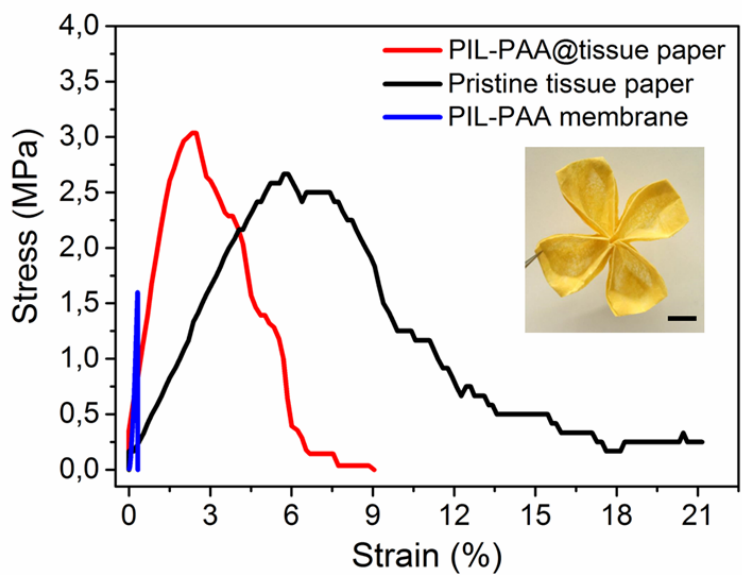

**Figure 3.** The mechanical testing plots of the hybrid PIL-PAA@tissue paper membrane (red), pristine tissue paper (black) and PIL-PAA membrane (blue). Inset is the photograph of hybrid membrane folded in a flower form, scale bar: 2 cm.

To investigate the actuation performance of the as-prepared hybrid PIL-PAA@tissue paper membrane, a membrane strip (20 mm × 1 mm × 100 μm) was loaded in an acetone vapor phase (24 kPa, 20 °C). Acetone was chosen as it was used as a reference solvent for the actuation study in our previous system.[27] The flat membrane was observed initially to bend from its flat state quickly into a loop with top surface inwards in 6 s (Figure 4a and 4b), resulting in a bending curvature of 0.314 mm$^{-1}$ (Figure S10). Here we also studied the effect of membrane thickness on the bending angle (Figure S11), and calculated 'curvature × membrane thickness' of our membrane at corresponding membrane thickness. Then we compare our membrane with previous literature results (Table S1). It is clear that our membrane shows a comparably good performance in terms of the response speed and the amplitude of locomotion. Upon exposure back to air, the membrane rolled slowly into a loop in an opposite direction with the top surface outwards (Figure 4a). After this step, the hybrid membrane undergoes actuation between these two positions of different bending directions. The detailed kinetics of the actuation movements is assessed by plotting the bending angle of the membrane against time by repeatedly transferring the membrane between acetone vapor and air. It is observed that the membrane completes each actuation cycle between two end positions within 15 s in a repetitive manner (Figure 4c). By contrast, the pristine tissue paper only slightly and slowly deformed its shape in response to acetone vapor because of a slight structural asymmetry generated during the tissue paper production. In the PIL-PAA@tissue paper hybrid membrane, asymmetric stress distribution in the membrane was built up due to the porous, gradient structure along the membrane cross-section created by ammonia treatment, and consequently acetone molecules swell the bottom more than the top part, thus bending the membrane, and *vise versa*.

Furthermore, several other papers including napkin paper, printing paper and filter paper were transformed into porous hybrid membranes. The napkin paper-based one shows a similar porous structure and morphology with tissue paper (Figure S12), and reaches a bending angle of 205$^o$ in acetone vapor in 6 s (Figure S13). However, in the case of printing paper or filter paper substrate, the porous network was poorly formed, showing a discontinuous PIL component incorporated into the cellulose fiber networks (Figure S14 and S15) and displaying no response to acetone vapor (Figure S13). When considering the mass density in area of these papers (Figure S13), they are much higher in the printing paper and filter paper than that in the tissue paper and napkin paper, indicating a denser packing of cellulose fibers inside the paper, *i.e.* a lower porosity, which is confirmed in the mercury porosimetry measurement by less mercury uptake by the printing paper (0.9 mL/g at 100 MPa in Figure S16) in comparison to the tissue paper (4.5 mL/g, at 100 MPa). Here, the mass density as well as the porosity matters for maintaining a homogeneous porous PIL network as the solution impregnation method relies on the interstice in cellulose fiber networks. Thus, a higher mass density and a lower porosity in printing paper limit the formation of the porous PIL network, which is the driving force for actuation.

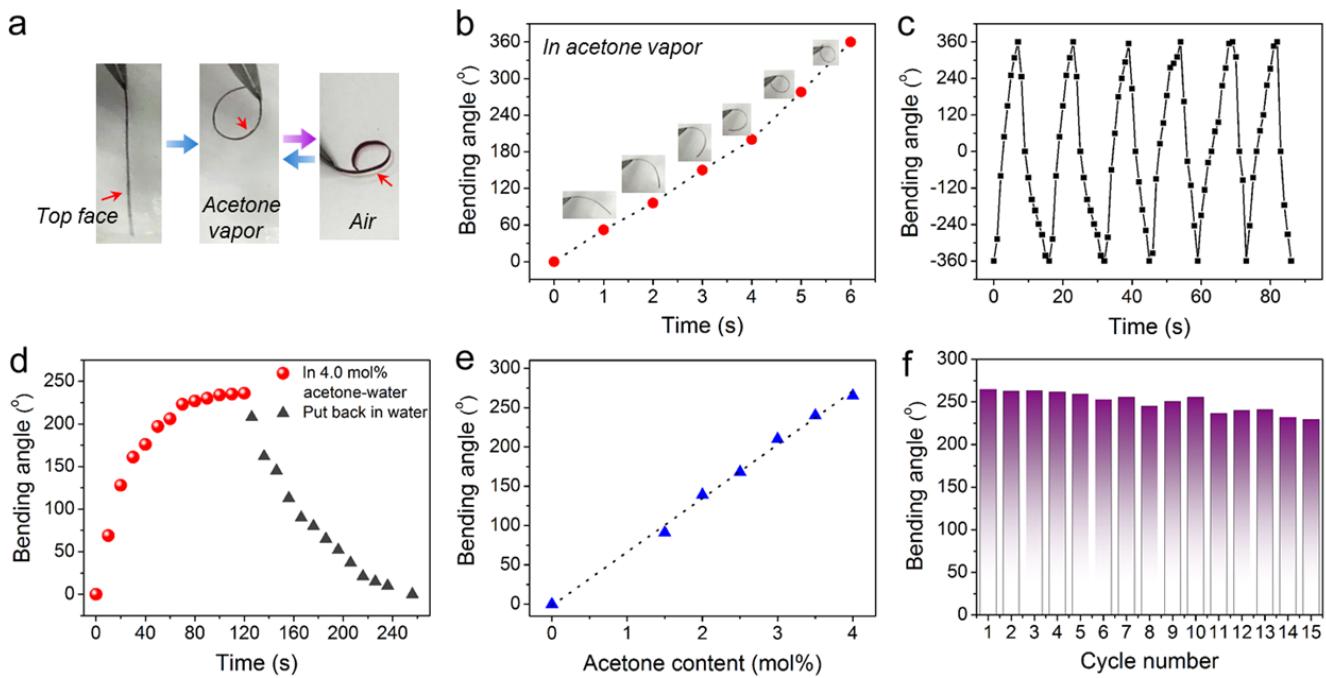

**Figure 4.** Actuation performance of the hybrid PIL-PAA@tissue paper membrane. (a) Photographs of the deformation of the membrane strip in acetone vapor (shown in blue arrow) and air (shown in purple arrow), red arrow represents the top face of hybrid membrane. (b) Dependence of the bending angle of a fresh flat membrane upon acetone vapor. The insets are the photographs of the deformation of the membrane strip in acetone vapor. (c) Dependence of the bending angle of the membrane upon air and acetone vapor under 6 cycles. (d) Dependence of the bending angle against time when placing it in a 4.0 mol% acetone–water mixture and back in water. (e) Dependence of the bending angle on acetone content (mol%). (f) Dependence of the bending angle on cycle number in 4.0 mol% acetone-water mixture.

The porous PIL-PAA@tissue paper membrane can not only respond to solvent vapor but also bend in the solvent-water phase. When the membrane was placed in 4.0 mol% acetone-water, it achieved a bending angle of 236° in 2 min, and then gradually recovered when stored in water (Figure 4d). By increasing acetone content from 1.5 to 4.0 mol%, the membrane bent inwards continuously and ended up with a maximum bending angle of 265° (Figure 4e). It shows that the bending angle of the membrane is linearly proportional to liquid acetone content. This continues bending–unbending cycle in 4.0 mol% acetone–water mixture can be repeated at least 15 times with high reversibility (Figure 4f). Moreover, the membrane can also respond to other solvents with different bending angle after 5 min (*e.g.*, tetrahydrofuran: 360°; methanol: 14°; isopropanol: 20°), showing the different sensitivity to solvents (Figure S17). The above results demonstrated the resultant hybrid membrane has potential application in sensing a wide variety of solvents.

Importantly, due to the advantageous mechanical flexibility, the hybrid membrane strips can be further woven into a self-supported textile structure (Figure S18). When wearing them with top surfaces on one side, this textile generated cooperative actuation bending with top surfaces inwards in contact with acetone vapor. When exposed to air, it completely recovered its flat state and continued to bend towards the opposite side. More excitingly, the hybrid membrane can realize, apart from simple bending, contractile/extensile actuators. The preparation of the membrane strips with an "**S**" shape is illustrated in Figure S19. A long membrane strip was cut into three shorter ones. The central strip was placed upside-down and joint with the other 2 strips at their cutting edges together *via* an epoxy resin. The glued membrane generated contractive **S**-shape actuation when exposed to acetone vapor (Figure 5a). Specifically, the contractile strain increased with the decreasing distance between the membrane strip and liquid acetone surface, following the increasing acetone vapor pressure. A maximum contractile strain of 82% was achieved at a distance of 0.8 cm above liquid acetone (Figure 5b). The mechanical force generated by the contracted membrane was measured by locating a contracted strip 0.2 cm above an electronic balance. With the diffusion out of the acetone molecule from a membrane strip, the contracted membrane stretched down and exerted a force on the balance, which reached a maximum of 0.47 mN (47 mg, Figure 5c). The sensitivity of the actuation was affected by the type of solvent vapors. For instance, slight contractile response was observed in the vapor phase of isopropanol and methanol, which can only swell but not dissolve PIL. In contrast, the strip exhibited much higher contractile strains in response to vapors of acetone and THF, as they can dissolve PIL, possessing lower boiling point and strong "solvent ~ PIL" interactions (Figure 5d). This result suggests that the membrane can serve as a potential sensor for different solvent vapors. Besides, the strip exhibited high contractile reversibility without significant decrease in contractile strain at a distance of 0.8 cm after 50 cycles (Figure 5e). The principle to build up the **S**-shaped contractor can be extended. For example, when cut into four short pieces instead of three and reconnected by glue with mismatched top-bottom surface, the strip produced a wave that is of even longer contraction distance (Figure S20).

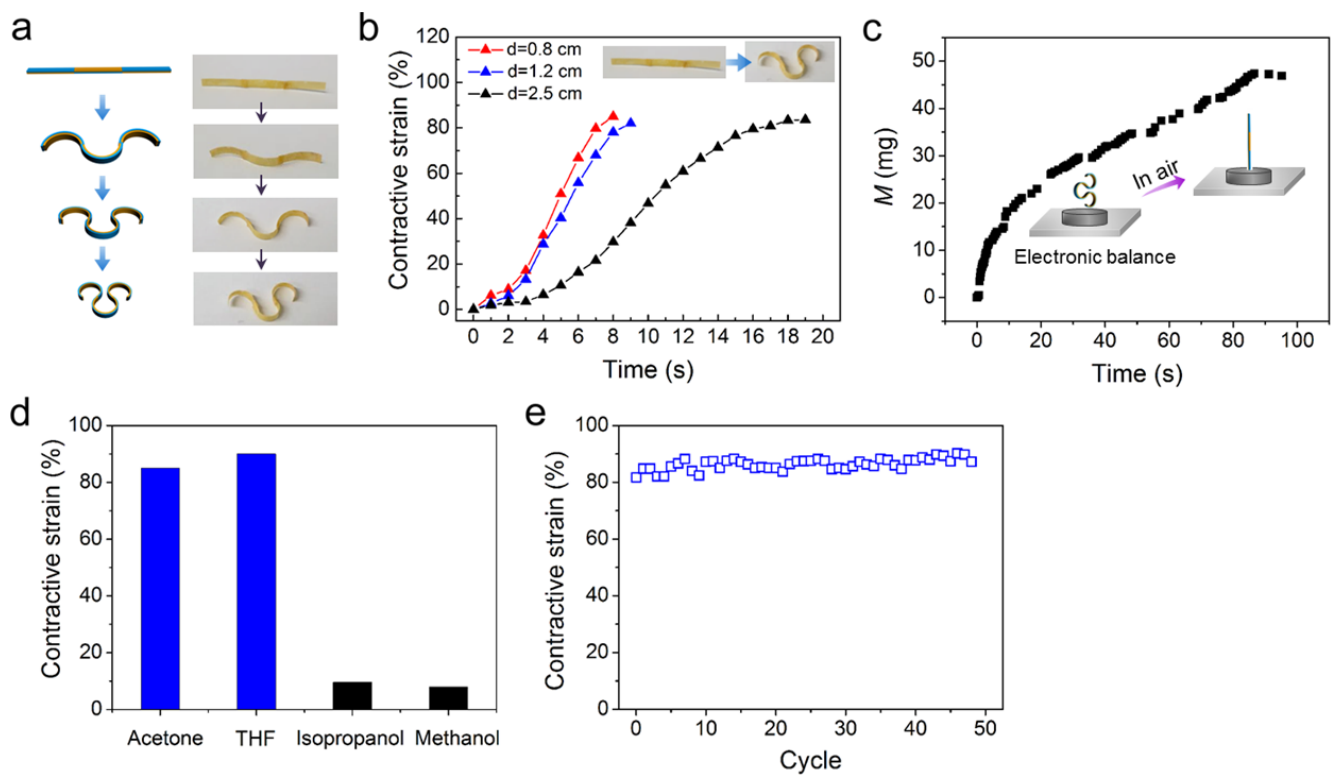

**Figure 5.** (a) Schematic illustration and photographs of the contractive actuation generated by the **S**-shape strip upon exposure to acetone vapor. (b) Dependence of the contractive strain on time at different distances between membrane strip and acetone liquid phase. The contractive strain was calculated from the ratio of the strip length to its original length. (c) Plot of the force ($M$) generated by contracted membrane actuator against time when exposed to air. Inset is the schematic illustration of the experimental force measurement set-up. (d) Dependence of the contractive strain on different solvent vapors: acetone, THF, isopropanol and methanol. (e) Dependence of contractive strain on cycle number (distance: 0.8 cm). The strip was put back into water for 2 min after every 5 cycles.

Importantly, to produce actuators of even higher complexity in motion[54], the **S**-shape strips are connected. For example, three pieces of them are connected together on mismatched top-bottom surfaces (top-to-bottom), top-top surfaces (top-to-top) or bottom-bottom surfaces (bottom-to-bottom) (Figure 6). The strip made from mismatched top-bottom surfaces generated a flower-like shape when exposed to acetone vapor (Figure 6a). Conversely, the strip made from top-to-top method with an outstretched state in air contracted to the center upon acetone vapor (Figure 6b). The strip made from bottom-to-bottom method showed a contracting state, and then started to extend out of the center when exposed to acetone vapor (Figure 6c). Therefore, complex motions of different forms can be designed from the same **S**-shape strip *via* different end matching.

Furthermore, these hybrid membranes can be widely used as a new sort of artificial materials. Herein, three strips of hybrid membranes (weight: 15 mg) were connected together on the same side to suspend a star-like load with a weight of 232 mg. When exposure to acetone vapor, the membranes contracted to lift up the object (Figure S21a). Also, the contractile actuation produced from the hybrid membranes can be designed to behave as a retractable hook to pull down a toy window shield (Figure S21b) or lift up an object by cooperative contraction from a long mismatched top-bottom membrane (Figure S21c). These examples presented here only depict limited actuation behavior. Considering the possible combination of membrane strips of different hierarchy, *i.e.* strips bearing different number of blocks and various connection models of individual strip (linear and branched), the accessible actuation forms can be eventually countless for a wide range of applications.

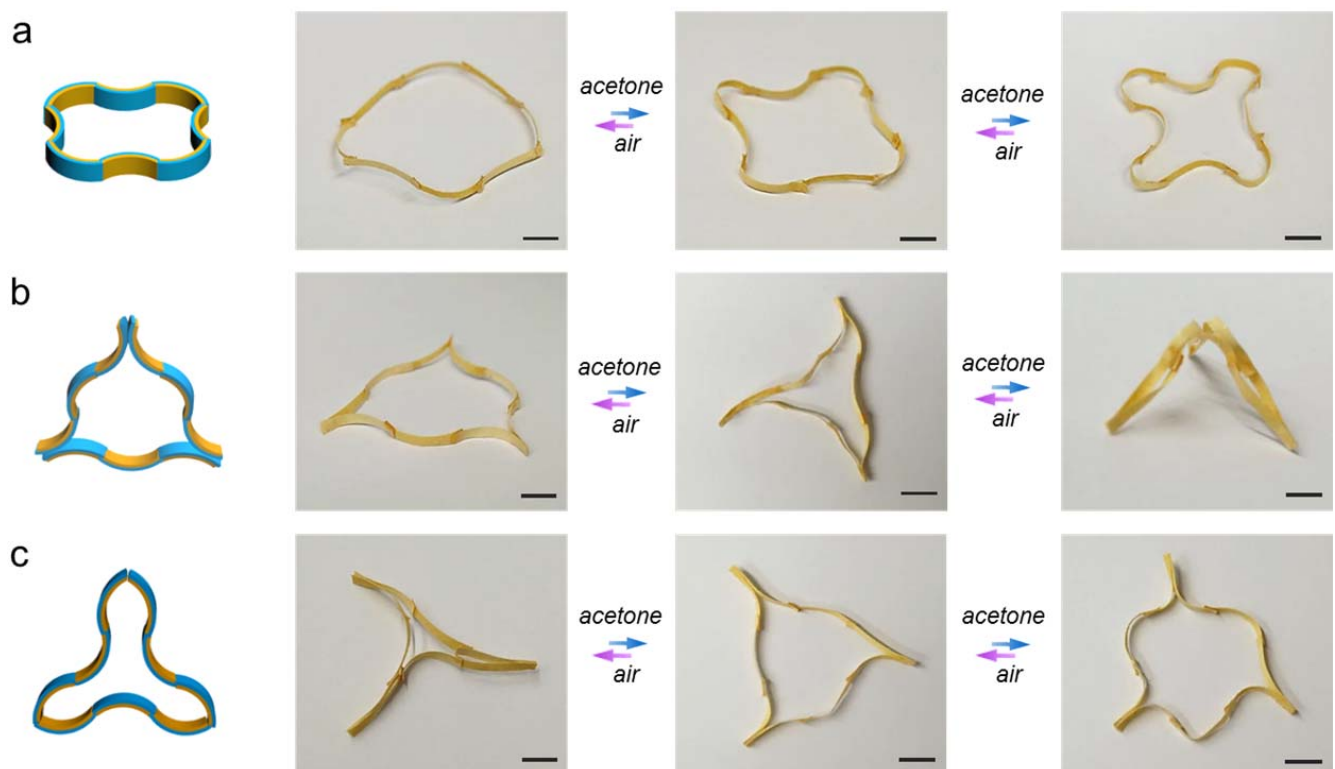

**Figure 6.** Schematic illustration and photographs of the connected S-shape strips on mismatched top-bottom surfaces (top-to-bottom), top-top surfaces (top-to-top) or bottom-bottom surfaces (bottom-to-bottom) bent in response to acetone vapor. Scale bar: 1 cm.

## 4. CONCLUSIONS

In summary, a flexible porous membrane was developed in a simple concept by incorporation of a nanoporous gradient poly(ionic liquid) network into a commercially available tissue paper. The mechanical flexibility derived from the tissue paper component due to long cellulose microfibers, and the nanoporous gradient structure of the membrane are synergistically combined to enable new textile actuator designs. As a proof of concept, **S**-shaped contractors and their assemblies exhibit physical motions far beyond the classical bending form of most thin-film-based actuators. This simple yet versatile fabrication concept opens up the development of new forms of actuators with targeted physical motions, which may have a wide window of applications including sensors for harmful chemicals, robotics, energy production, *etc*.

SUPPORTING INFORMATION

Additional SEM image, EDX, FT-IR characterization and actuation data. The Supporting Information is available free of charge via the Internet at http://pubs.acs.org.

AUTHOR INFORMATION

**Corresponding Authors**

*E-Mail: jiayin.yuan@mpikg.mpg.de;

*E-Mail: john.dunlop@mpikg.mpg.de

**Notes**

The authors declare no competing financial interest.

ACKNOWLEDGMENT

This work was supported by the International Max Planck Research School (IMPRS) on "Multiscale Biosystems".

# SUPPORTING INFORMATION

# Flexible and Actuating Nanoporous Poly(ionic liquids)-paper based Hybrid Membranes

*Huijuan Lin,[†] Jiang Gong,[†] Han Miao,[†] Ryan Guterman,[†] Haojie Song,[‡] Qiang Zhao,[†] John W. C. Dunlop[§],\* and Jiayin Yuan[†],\**

[†]Department of Colloid Chemistry, Max Planck Institute of Colloids and Interfaces,

Am Mühlenberg 1 OT Golm, D-14476 Potsdam, Germany

[‡]School of Materials Science and Engineering, Jiangsu University,

Zhenjiang, Jiangsu, 212013, China

[§]Department of Biomaterials, Max Planck Institute of Colloids and Interfaces,

Am Mühlenberg 1 OT Golm, D-14476 Potsdam, Germany

\*To whom correspondence should be addressed.

\*E-Mail: jiayin.yuan@mpikg.mpg.de; \*E-Mail: john.dunlop@mpikg.mpg.de

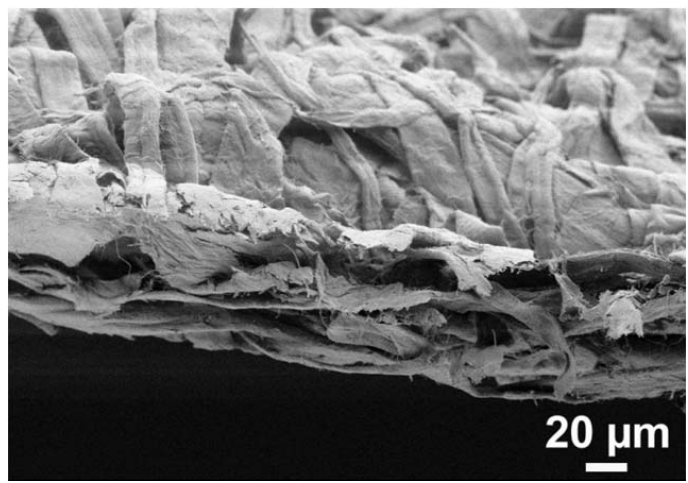

**Figure S1.** SEM image of the cross-section view of a pristine tissue paper.

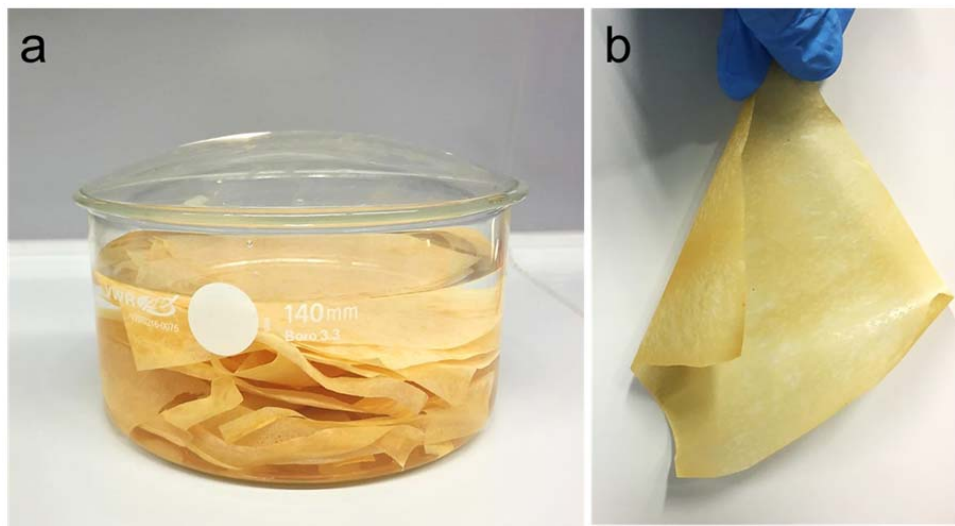

**Figure S2.** (a) Photograph of six pieces of free-standing PIL-PAA@tissue paper membranes (overall size = 480 cm$^2$) stored in water. (b) Photograph of one PIL-PAA@tissue paper membrane with a size of 10 cm × 8 cm.

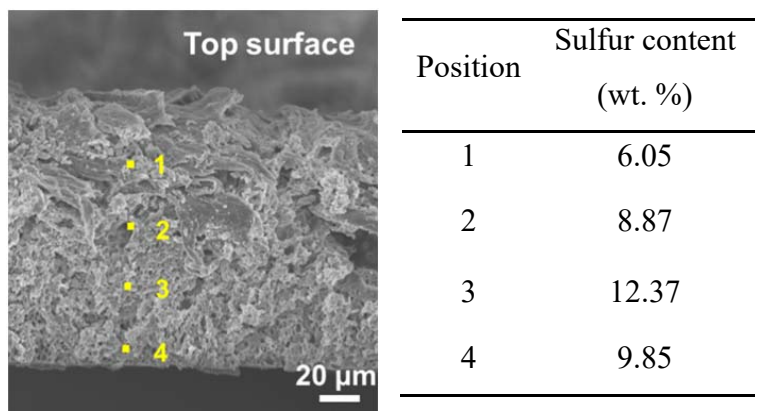

**Figure S3.** Sulfur content from top to bottom along the cross-section of the hybrid membrane. Inset is the cross-sectional SEM image of the hybrid membrane and the positions (1-4) taken for EDX spectra analysis. Position 4 is unexpectedly lower, possibly due to back-diffusion of aqueous ammonia into the polymer blend/paper film on the PTFE plate.

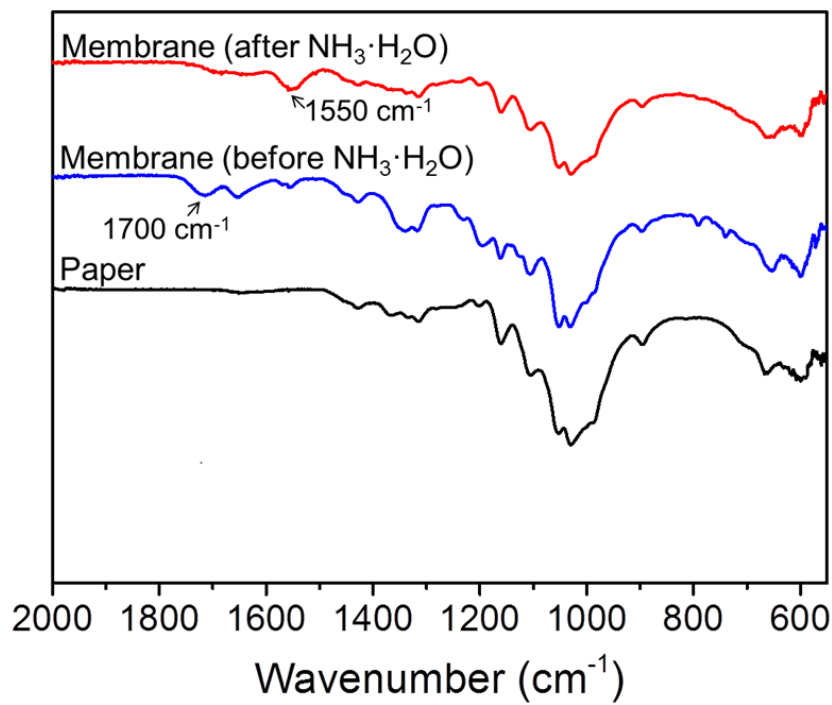

**Figure S4.** FT-IR spectra of the hybrid PIL-PAA@tissue paper membrane (before and after $NH_3 \cdot H_2O$ treatment) and the pristine tissue paper.

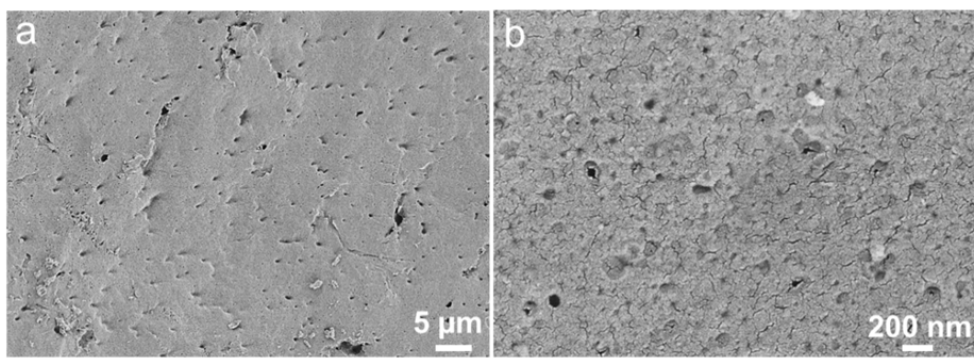

**Figure S5.** SEM images of the hybrid PIL-PAA@tissue paper membrane in a bottom view in a low (a) and high (b) magnification, respectively.

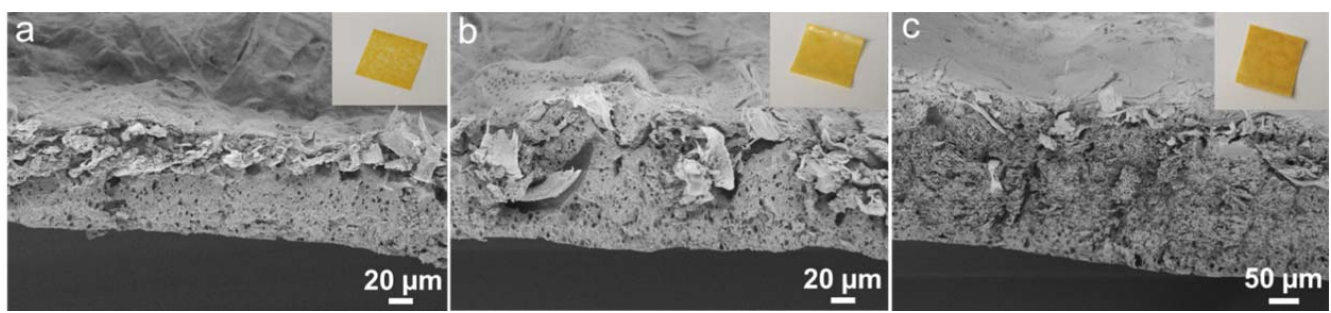

**Figure S6.** SEM images of the PIL-PAA@tissue paper membranes (cross-section view) with a coating volume of PIL-PAA solution ranging from 0.6, 1.0 and 1.4 mL, respectively (concentration of PIL: 100 mg/mL). Insets are the photographs of the hybrid membranes (size: 2 cm × 2 cm). The thickness of the membranes is approximately 85, 115 and 240 μm, respectively.

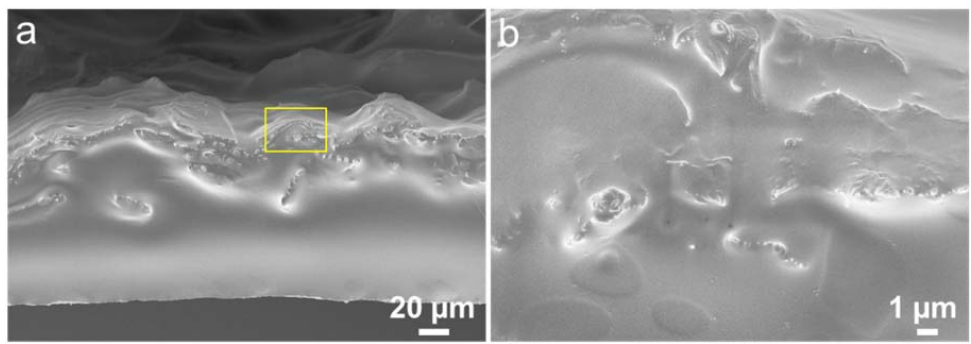

**Figure S7.** Cross-sectional SEM images of the hybrid PIL-PAA@tissue paper membrane before ammonia treatment.

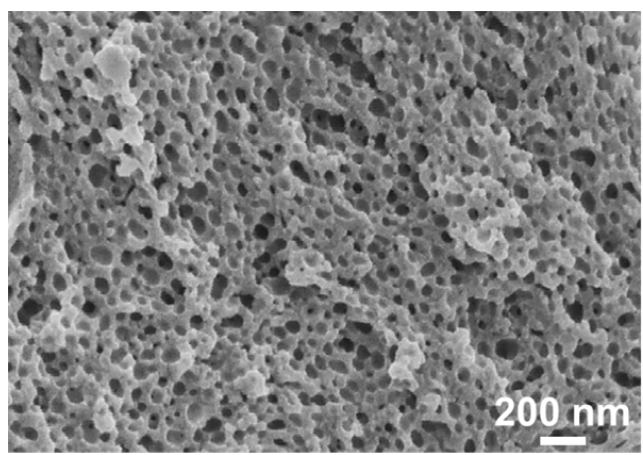

**Figure S8.** SEM image of a tissue paper-free porous PIL-PAA membrane in a cross-section view.

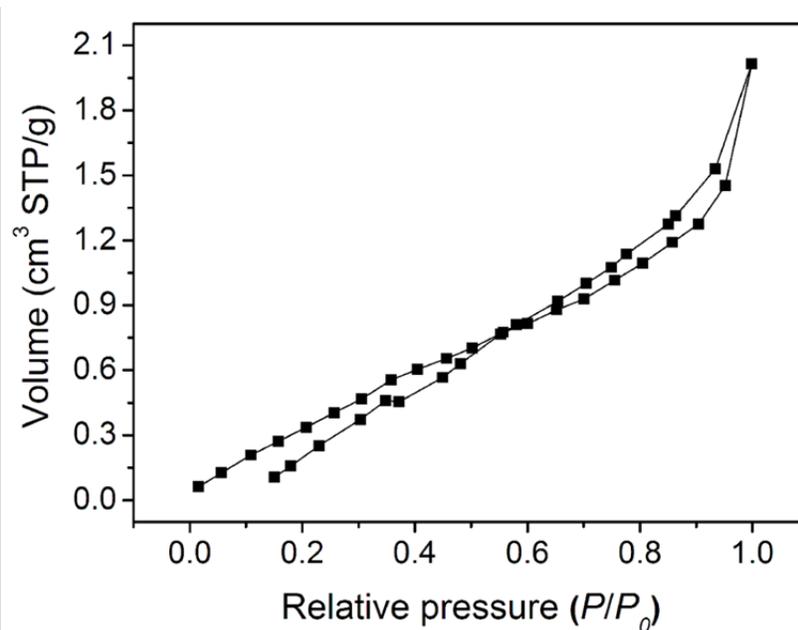

**Figure S9.** Nitrogen adsorption/desorption isotherms of the PIL-PAA@tissue paper membrane.

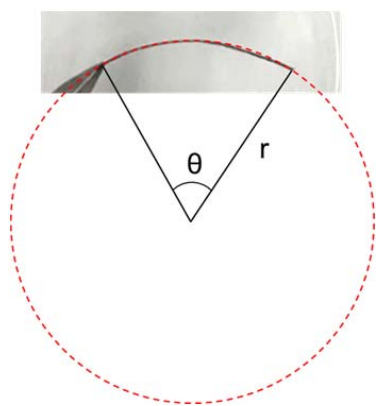

$$L = \frac{2\pi r \theta}{360}$$

θ: central angle (degree)

r: radius (mm)

L: length of the actuator membrane (20 mm)

**Figure S10.** Measuring the bending angle or central angle (θ, degree) of the PIL-PAA@tissue paper membrane (20 mm × 1mm × 100 μm) in acetone vapor (24 kPa, 20 °C). The curvature for comparison in Table S1 was calculated based on the following equation: Curvature $= \frac{1}{r}$. When the membrane bends into a close loop (θ = 360°), the bending curvature will be: Curvature $= \frac{1}{r} = 0.314$ (mm$^{-1}$).

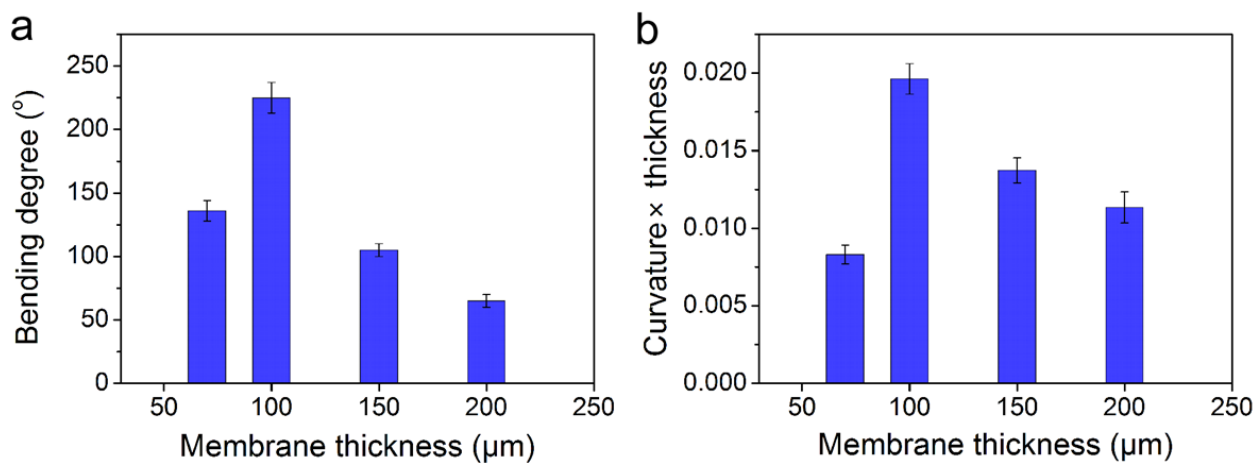

**Figure S11.** (a) Dependence of bending degree on the thickness of the PIL-PAA/tissue paper membrane at the time of 5$^{th}$ second. (b) Dependence of the normalized curvature (curvature × thickness) on the thickness of the PIL-PAA/tissue paper membrane.

**Table S1.** The bending performance in this work and the previous literature results. Note: the actuators that bend in one way are considered. In each cited paper, we choose the fastest speed for comparison.

| Actuator thickness (mm) | Curvature (mm$^{-1}$) | Curvature × thickness | Time (s) | Ref. in *SI* |
|---|---|---|---|---|
| 0.1 | 0.314 | 0.0314 | 6 | This work |
| 0.003 | 0.48544 | 0.00146 | 56 | S1 |
| 0.029 | 0.1428 | 0.00414 | 15 | S2 |
| 0.056 | 0.15924 | 0.00892 | 120 | S3 |
| 0.5 | 0.5181 | 0.02591 | 420 | S4 |
| 0.127 | 0.0813 | 0.01033 | 30 | S5 |
| 0.05 | 0.02 | 0.001 | 50 | S6 |
| 0.03 | 0.49505 | 0.01485 | 20 | S7 |
| 0.12 | 0.26964 | 0.02516 | 20 | S8 |

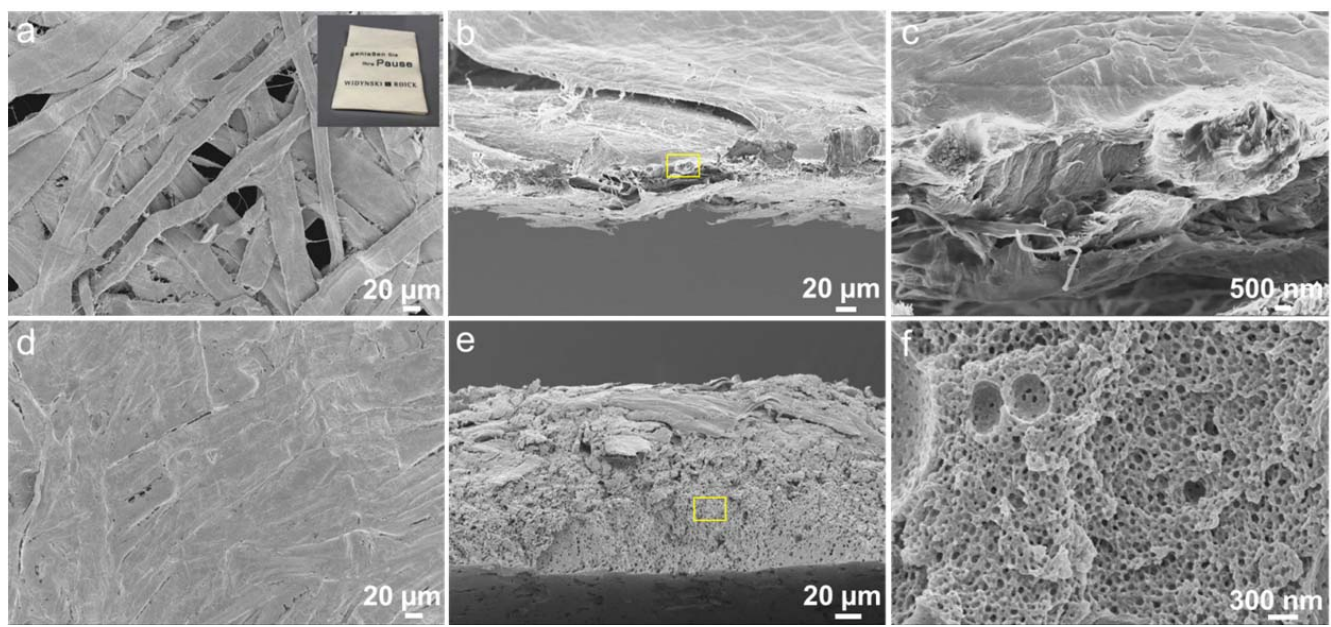

**Figure S12.** (a) SEM image of the top view of a napkin paper. (b) SEM image of the cross-section view. (c) SEM image of the region marked with a yellow box in (b). (d) SEM image of the top view of the PIL-PAA@napkin paper membrane. (e) SEM image of the cross-section view. (f) SEM image of the region marked with a yellow box in (e).

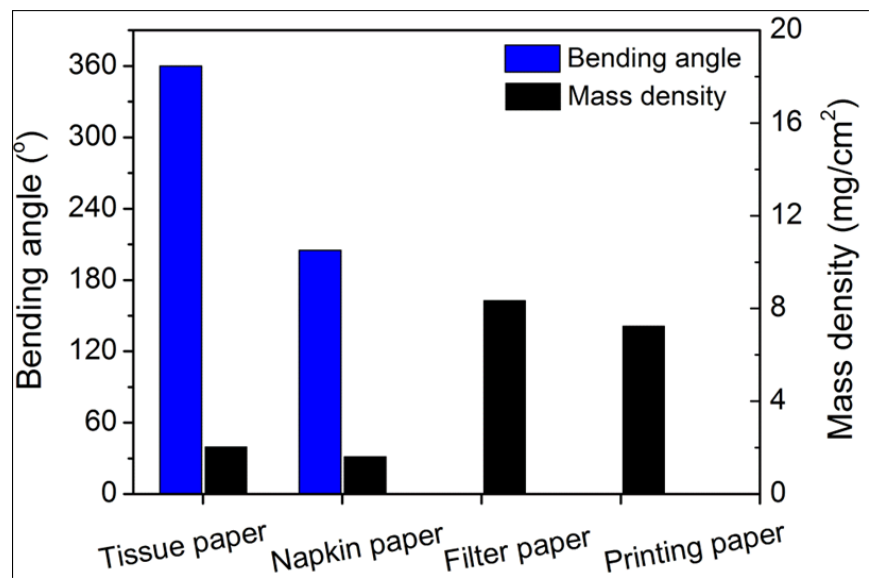

**Figure S13.** The bending angle of the prepared membrane from different paper substrates at 6 s in acetone vapor and the mass density of the paper substrates.

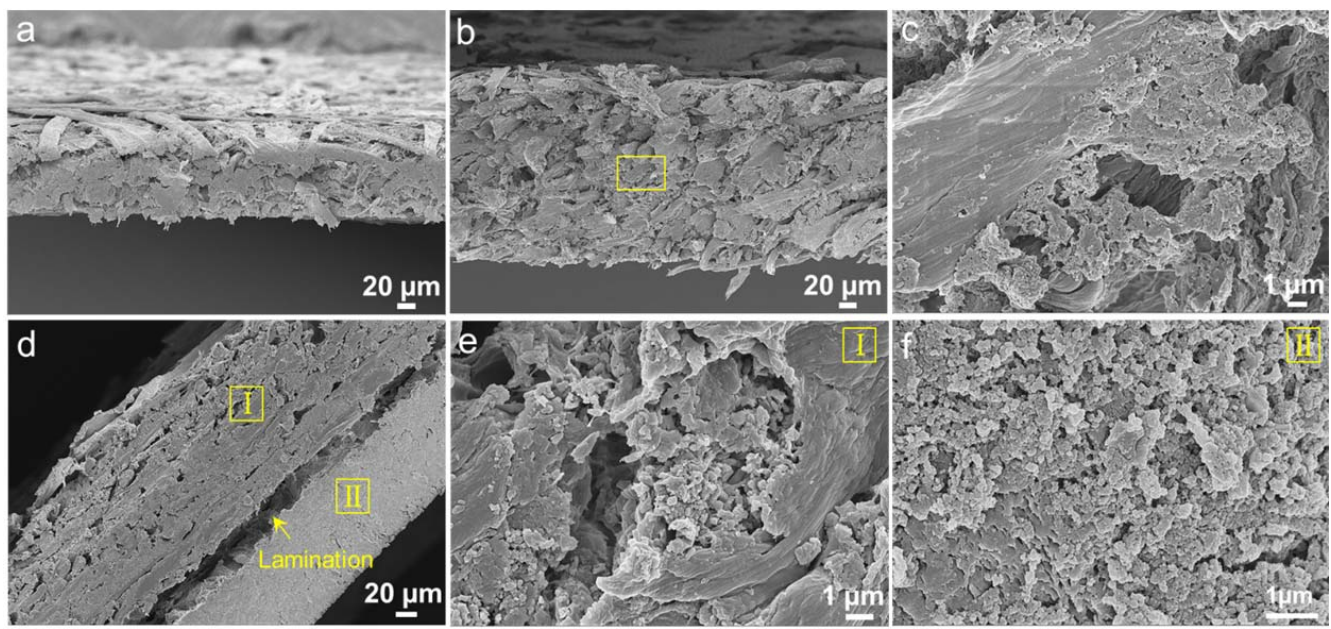

**Figure S14.** SEM images of the cross-section view of a printing paper (a) and a PIL-PAA@printing paper membrane in a low (b) and high (c) magnification. When the coating volume of the polymer solution was increased (d-f), the delamination was formed obviously.

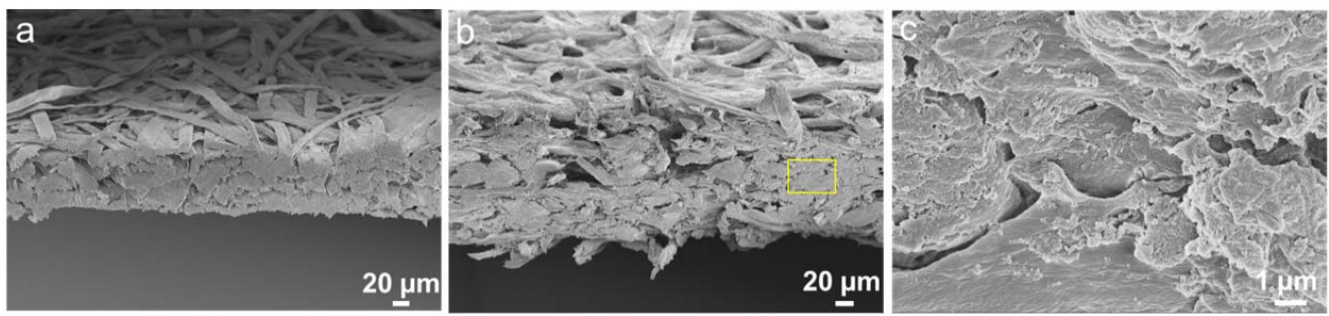

**Figure S15.** SEM images of the cross-section view of a filter paper (a) and the PIL-PAA@filter paper membrane in a low (b) and high (c) magnification.

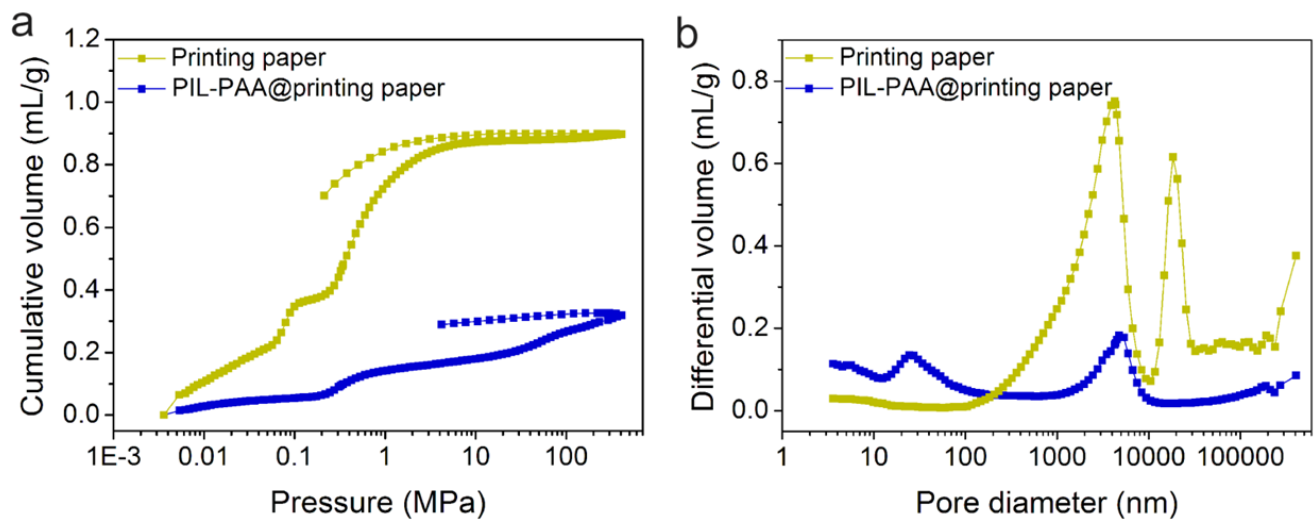

**Figure S16.** Porosity and pore size distribution of printing paper and PIL-PAA@printing paper by mercury intrusion measurement. (a) Cumulative mercury volume as a function of the pressure. (b) The differential pore volume as a function of the pore size. The porosity of printing paper and PIL-PAA@printing paper is 62% and 33%, respectively.

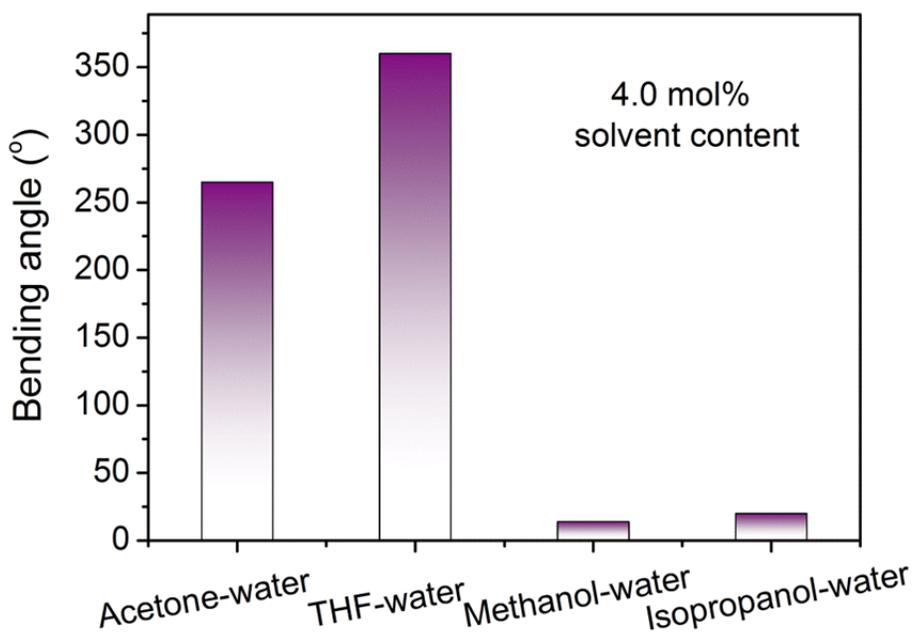

**Figure S17.** Dependence of the bending angle of the PIL-PAA@tissue paper on solvent types (solvent content: 4.0 mol%).

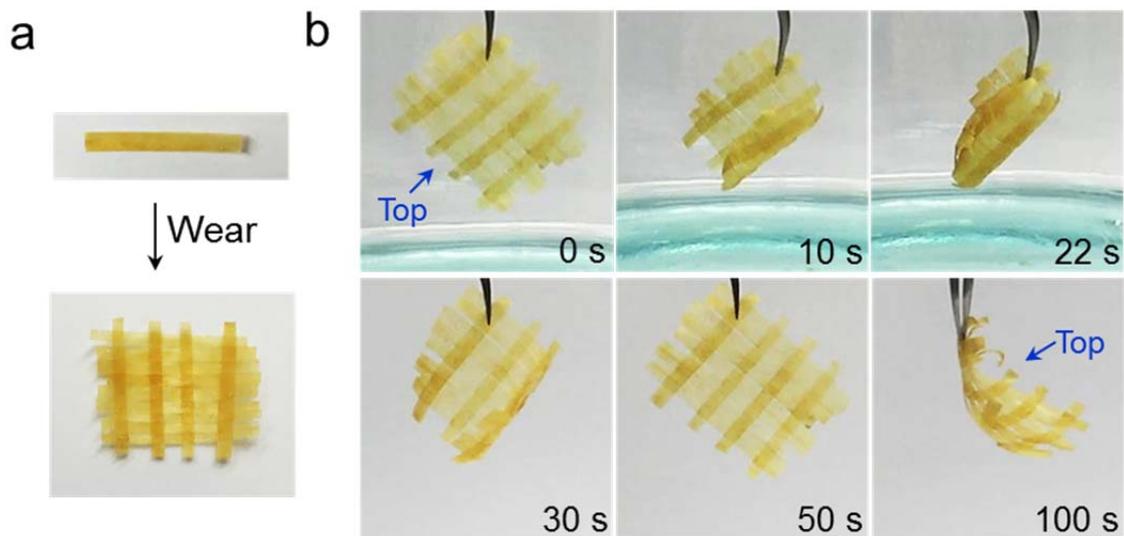

**Figure S18.** (a) A smart textile (size: 2.5 cm × 2 cm) woven from the hybrid PIL-PAA@tissue paper membrane strips. (b) Actuation of the responsive textile in acetone vapor and air.

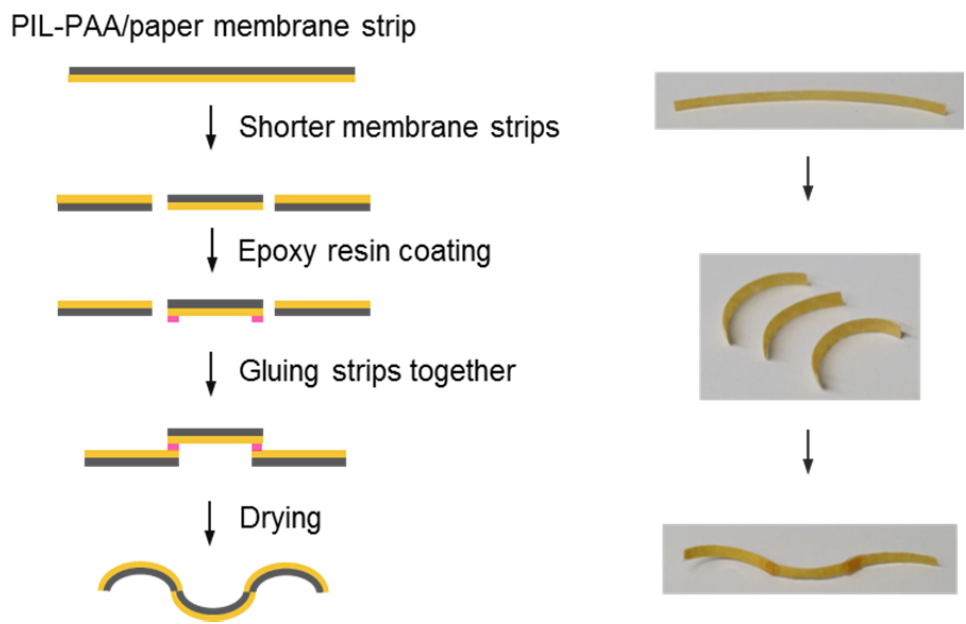

**Figure S19.** Assembly process of the **S**-shape membrane strip. Three pieces of short strips were cut from a long membrane strip and then glued together at the cutting position. The yellow and grey colors represent the top and bottom surfaces of the membrane, respectively. The rose red is to represent epoxy resin.

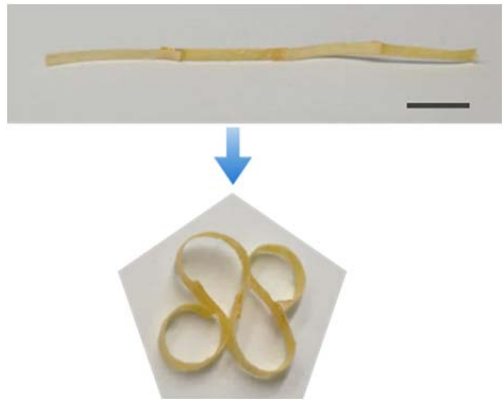

**Figure S20.** Deformation of a wave-shape membrane strip in acetone vapor. The actuator was prepared from a long membrane strip, which was cut into 4 pieces and reconnected by mismatch of the top-bottom surface of adjacent strips. Scale bar: 1 cm.

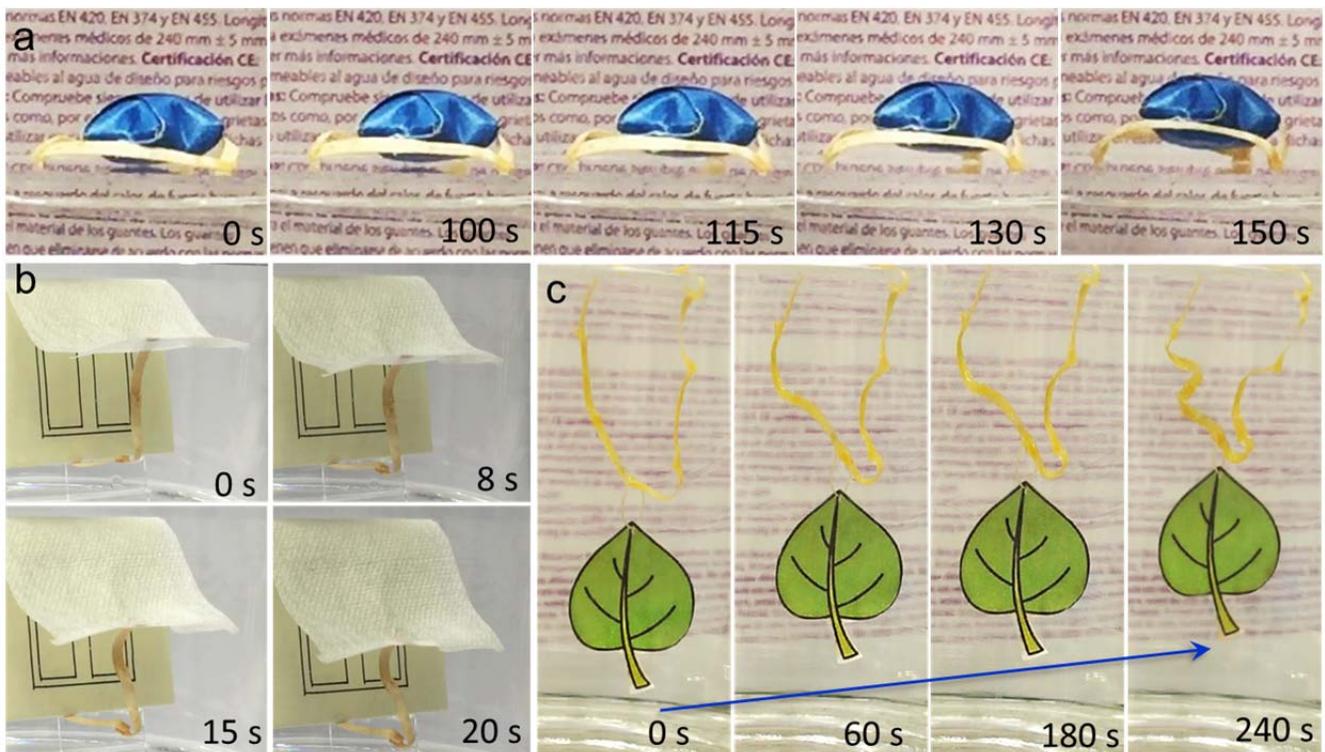

**Figure S21.** (a) Three membrane strips connected together (weight: 15 mg) to lift up a star-like object in acetone vapor, 50 °C. The weight of the object is 232 mg. (b) **S**-shape membrane pulled down a toy window shield due to the contractile actuation in acetone vapor, 20 °C. (c) An object was lifted up by cooperative contraction driven by acetone vapor, 20 °C. The weight of the object is 25.9 mg.